\documentclass[11pt]{article}
\usepackage[utf8]{inputenc}
\usepackage[T1]{fontenc}
\usepackage{lmodern}
\usepackage{microtype}
\usepackage{graphicx}
\usepackage{hyperref}
\usepackage{booktabs}
\usepackage{amsmath}
\usepackage{tabularx}
\usepackage{tikz}
\usetikzlibrary{arrows.meta,positioning,fit,backgrounds,calc,shapes.geometric}
\usepackage{url}
\usepackage[numbers,sort&compress]{natbib}

\title{Toward Self-Evolution-Ready Workflow Harnesses:\\
  A Reversible Migration Path and Convertibility Taxonomy\\
  for Expert LLM Pipelines}
\author{
  Yimo Lin\textsuperscript{1}\thanks{Corresponding author: \texttt{murdmonoto@gmail.com}} \and
  Zhen Zhang\textsuperscript{2} \and
  Yibin Li\textsuperscript{1} \\[2pt]
  \textsuperscript{1}Yunyong Century (Beijing) Artificial Intelligence Technology Co., Ltd. \\
  \textsuperscript{2}Tsinghua University
}
\date{2026}

\begin{document}
\maketitle

\begin{abstract}
Expert-validated ``LLM + script'' workflows create real value but remain frozen: they encode hard-won judgment yet cannot adapt from their own results.

Existing agent research largely targets greenfield agents and benchmark settings; it does not report how to migrate a running, expert-validated legacy workflow into a system that is ready to adapt.

We present a reversible, Strangler-Fig migration path that turns such a workflow into composable, typed, auditable stages, and a convertibility taxonomy (A/B/C)---realized as a routing stage in the harness---that diagnoses whether a given workflow can travel this path and routes it accordingly.

On a production WeChat Official Account workflow, 9 expert functions became 9 independently runnable tools, with 0 business-logic changes, one-flag rollback, and fully traceable execution under a small set of deterministic safety invariants.

We report an early, strategy-level outcome-feedback signal (autonomous topic-selection tuning) as supporting evidence of self-evolution readiness---not a validated self-learning result. The contribution is the migration path and taxonomy that make legacy workflows self-evolution-ready, honestly bounded by a single-case study.

\end{abstract}
\section{Introduction}
As coding agents proliferate, many business processes have been validated by domain experts as ``LLM + script'' workflows. These workflows embed expert judgment but are frozen: a script that produces content cannot learn from whether that content succeeded.

The gap we address is migration, not invention: how does one incrementally turn an already-running, expert-validated legacy workflow into a harness that is ready to adapt---reversibly, auditably, and without disrupting the business it already runs? Recent mechanism-layer work targets greenfield designs or abstract evaluation; none reports the lived migration of a real legacy business workflow toward an adaptive system.

Our central claim is deliberately narrow. We do not claim to have built a fully self-evolving agent. We claim a migration methodology that makes legacy expert workflows self-evolution-ready, plus a diagnostic taxonomy for when that migration is feasible. The strongest evidence in this paper is migration cost, stage decomposition, reversibility, and auditability; the self-evolution signal is reported as early, supporting evidence only.

\subsection{Contributions}
\begin{itemize}
  \item C1 (primary) --- A reversible Strangler-Fig~\cite{stranglerfig} migration path that decomposes a legacy expert script into composable, independently runnable stages, with zero business-logic change and one-flag rollback at every step.
  \item C2 (primary) --- A convertibility taxonomy (A/B/C), realized as a routing stage in the harness, with a diagnostic checklist and migration-cost estimates that determine whether a workflow can be decomposed into stages or must first be refactored---and that route it accordingly.
  \item C3 (supporting evidence) --- An early, strategy-level self-evolution signal: the deployed system autonomously tunes topic selection from outcome review. Reported as a first production data point, not a validated learning result.
  \item C4 (guardrail / engineering requirement) --- A deliberately minimal deterministic safety check that makes migration into a live business permissible. Safety is not the contribution; safety is what makes migration permissible.
\end{itemize}
\section{Background and Problem}
We define four notions used throughout. A frozen expert workflow is a stable script encoding expert judgment with no feedback from outcomes. A self-evolution-ready harness is a runtime whose pipeline is decomposed into composable stages and instrumented so that outcomes can, in principle, drive changes to strategy or composition. A stage contract is a stage's typed input/output plus its declared function, making it addressable by what it does. Convertibility is whether a given workflow can be decomposed into such stages, or must be refactored first.

To keep claims precise, we use a self-evolution maturity ladder. It lets us state exactly how far this paper reaches without overclaiming: we complete L1, give early evidence of L2, and place L3--L5 as trajectory.

\begin{table}[htbp]\centering\footnotesize
\caption{Self-evolution maturity ladder.}
\begin{tabularx}{\textwidth}{l X X X}
\toprule
\textbf{Level} & \textbf{Name} & \textbf{Meaning} & \textbf{This paper} \\
\midrule
L0 & Frozen script & Fixed pipeline, no feedback & legacy baseline \\
L1 & Traceable harness & Composable, auditable, reversible & completed \\
L2 & Strategy tuning & Adjust topic/strategy from outcomes & early signal \\
L3 & Stage re-implementation & Swap a stage's implementation from outcomes & future \\
L4 & Cross-vertical recomposition & Recompose stages into a new vertical & ongoing \\
L5 & Autonomous business loop & Discover, execute, review, expand autonomously & trajectory \\
\bottomrule
\end{tabularx}
\end{table}
\section{Migration Method}
The migration follows the Strangler Fig pattern: wrap the legacy system, then incrementally replace it, keeping it live and reversible throughout. We apply it in five steps, each independently runnable and rollback-able:

\begin{itemize}
  \item (1) Subprocess wrap --- run the whole expert script as a black-box subprocess; 0 business-logic change.
  \item (2) Toolification --- expose each callable function as a tool with an I/O schema.
  \item (3) Stage composition --- compose tools into typed, contracted stages forming a declarative recipe.
  \item (4) Agent-driven decisions --- replace a hardcoded decision loop (e.g. rewrite/accept) with a structured-decision agent loop, adjudicated by deterministic code.
  \item (5) Rollback and audit --- keep both engines (subprocess/agent) behind a one-flag switch; every step emits a structured trace.
\end{itemize}
\section{Architecture}
The harness comprises: Stage (typed pipeline step, addressable by function), Tool (independently runnable capability implementing a stage, with I/O schema), Recipe (declarative composition of stages), Agent Loop (structured ReAct~\cite{react}-style decision engine), Safety Gate (small set of hardcoded invariants), Human Checkpoint (approval queue for irreversible steps), and Trace (structured audit log). A frozen legacy script is the degenerate case: one fixed recipe, no feedback.

Within a judgment stage, low-stakes choices are delegated to the LLM (which candidate, whether to rewrite, accept/give-up), while irreversible or costly actions are enforced by a small deterministic gate (draft-only; score $\geq$ 77; single account; resource caps) that the LLM cannot override---decision authority sits with the LLM, adjudication authority with the harness. This gate is deliberately minimal (a handful of comparisons); its role is a supporting guardrail (C4), not the focus of the paper, and keeping the trusted base small is what makes it safe to let the surrounding stages evolve.

Self-evolution loop (early). On top of the per-run loop sits a slower outer loop: published artifacts accrue outcomes (engagement, revenue); a review stage extracts content--performance regularities; the result updates the topic-selection stage. Autonomy is stratified by stakes: reversible strategy tuning runs autonomously; high-cost directional changes (switching lanes) pass a human checkpoint designed to recede. Because every step is a contracted stage writing to the trace, one can reconstruct which outcome changed which strategy.

Convertibility routing as a stage. Crucially, deciding whether a new workflow can enter the harness at all is not an external, one-off human study---it is itself a stage. A convertibility-routing stage takes a workflow description as input and emits a classification (Type A/B/C) plus a routing decision: Type A enters the mechanical four-step migration; Type B is routed to prompt-extraction first; Type C is held for Code-First refactoring. Modeling intake as a stage is what makes the harness extensible by composition rather than by bespoke engineering: the same contracted-stage discipline that lets the pipeline self-evolve also lets it screen and admit new verticals. We implement this routing as an automated stage: it takes a workflow description and emits the A/B/C classification and routing decision automatically (criteria in Section 7). Systematic evaluation of the classifier's accuracy across many intakes is ongoing.

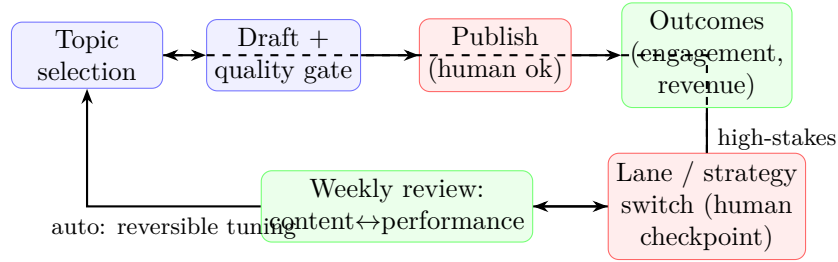
\begin{figure}[htbp]
\centering
\begin{tikzpicture}[
  font=\small,
  st/.style={draw, rounded corners, align=center, minimum height=7mm,
             minimum width=20mm, inner sep=3pt, fill=blue!6, draw=blue!55},
  rev/.style={draw, rounded corners, align=center, minimum height=7mm,
              minimum width=20mm, inner sep=3pt, fill=green!8, draw=green!55},
  gate/.style={draw, rounded corners, align=center, minimum height=7mm,
               inner sep=3pt, fill=red!7, draw=red!60},
  arr/.style={-{Stealth[length=2mm]}, thick},
]
\node[st] (topic) at (0,0)      {Topic\\selection};
\node[st] (draft) at (2.6,0)    {Draft +\\quality gate};
\node[gate](pub)   at (5.4,0)   {Publish\\(human ok)};
\node[rev](data)   at (8.2,0)   {Outcomes\\(engagement,\\revenue)};
\draw[arr] (topic)--(draft);
\draw[arr] (draft)--(pub);
\draw[arr] (pub)--(data);
\node[rev](review) at (4.1,-2.0) {Weekly review:\\content$\leftrightarrow$performance};
\draw[arr] (data) |- (review);
\draw[arr] (review) -| node[below,pos=0.25]{\footnotesize auto: reversible tuning} (topic);
\node[gate](dir) at (8.2,-2.0) {Lane / strategy\\switch (human\\checkpoint)};
\draw[arr] (review)--(dir);
\draw[arr,dashed] (dir) -- ++(0,1.0) node[right,pos=0.6]{\footnotesize high-stakes} |- (topic.east);
\end{tikzpicture}
\caption{The self-evolution loop. Inner loop produces and publishes; the outer
weekly review feeds outcomes back to topic selection. Reversible strategy
tuning is autonomous (solid); high-stakes directional switches pass a human
checkpoint (dashed) designed to recede. Publishing remains human-approved as a
routine, low-complexity step---not the locus of difficulty.}
\label{fig:loop}
\end{figure}

\section{Case Study: WeChat Content Workflow}
The migrated workflow collects high-engagement articles from benchmark accounts, AI-rewrites them under a content persona and writing rules, applies a ten-dimension quality scorer (hard threshold 77/100), generates a cover image, and saves a draft---never publishes. The expert system is a hardcoded sequential loop (auto\_publish.py, 917 lines; ai\_writer.py, 370 lines). It is Type A: nine functions mapped one-to-one onto nine tools, composed into stages, with zero business-logic change (only API-key literals moved to environment reads).

\begin{table}[htbp]\centering\footnotesize
\caption{Migration cost.}
\begin{tabularx}{\textwidth}{l X X}
\toprule
\textbf{Metric} & \textbf{Legacy script} & \textbf{Harness migration} \\
\midrule
Business-logic LOC changed & 0 & 0 \\
Expert functions wrapped & -- & 9 \\
Platform LOC & -- & 1,215 \\
Recipe / config LOC & -- & 1,340 \\
Test LOC & -- & 805 \\
Rollback mechanism & manual & one-flag engine switch \\
Production disruption & none & none \\
\bottomrule
\end{tabularx}
\end{table}
\begin{table}[htbp]\centering\footnotesize
\caption{Stage decomposition (contracts).}
\begin{tabularx}{\textwidth}{l X X X X X}
\toprule
\textbf{Stage} & \textbf{Legacy fn} & \textbf{Input schema} & \textbf{Output schema} & \textbf{Reusable?} & \textbf{Safety gate} \\
\midrule
candidate collection & list\_candidates & account, num & candidate[] & yes & account gate \\
rewriting & rewrite & article, persona & draft & yes & rewrite cap \\
quality scoring & quality\_check & draft & score, issues & yes & score $\geq$ 77 \\
image generation & gen\_images & topic, draft & image assets & partial & resource cap \\
draft save & save\_draft & draft, assets & draft\_id & yes & draft-only \\
\bottomrule
\end{tabularx}
\end{table}
A real run (draft \#100000026): first rewrite scored 76 (below threshold); the LLM judged it fixable with one rewrite remaining, rewrote, scored 83, judged it acceptable, and accepted. The subprocess version (\#100000014) uses a hardcoded ``rewrite at most once'' rule with no rationale. The same pipeline was validated on a financial-education account via a new config YAML, a writing-rules file, and an added compliance gate; platform layer and agent loop: zero changes.

\section{Evaluation}
\noindent\textbf{6.1 Migration cost \& reversibility.} Quantified in Table 2: zero business-logic change, nine tools, one-flag rollback, no production disruption. The subprocess and agent engines coexist permanently, so the legacy system is always a live fallback.

\noindent\textbf{6.2 Auditability.} Every LLM decision, tool call, gate trigger, and human checkpoint produces a structured trace event; secret-bearing fields are auto-redacted; a complete decision chain is reconstructible from the trace alone---which is also what makes an outcome-driven feedback loop auditable.

\noindent\textbf{6.3 Safety invariants.} We turn the ``9/9 adversarial tests'' into a complete safety case (Table 4). On our adversarial suite, boundary violations are intercepted at 100\% with 0 invariant violations; systematic per-invariant boundary-value coverage is future work. The result is unsurprising---a deterministic comparison is hard to fool---which is precisely the design intent.

\begin{table}[htbp]\centering\footnotesize
\caption{Safety invariants as a safety case.}
\begin{tabularx}{\textwidth}{l X X X}
\toprule
\textbf{Invariant} & \textbf{Risk prevented} & \textbf{Enforcement layer} & \textbf{Test} \\
\midrule
draft-only & unintended publication & deterministic code & static scan \\
score $\geq$ 77 & low-quality output & decision handler & adversarial accept test \\
API failure $\neq$ pass & false success & decision handler & fault injection \\
rewrite cap & runaway cost & runtime cap & adversarial loop test \\
single account & account contamination & config gate & account-mismatch test \\
\bottomrule
\end{tabularx}
\end{table}
\noindent\textbf{6.4 Subprocess vs agent engine.} Table 5 contrasts the two engines. The value of the agent engine is not ``smarter'' output but explainability, adjustability, and evolvability; we make no statistically significant quality claim from a single run.

\begin{table}[htbp]\centering\footnotesize
\caption{Subprocess engine vs agent engine.}
\begin{tabularx}{\textwidth}{l X X}
\toprule
\textbf{Dimension} & \textbf{Subprocess engine} & \textbf{Agent engine} \\
\midrule
Decision rule & hardcoded & LLM decision + deterministic adjudication \\
Rationale & none & structured natural-language rationale \\
Rollback & yes & yes \\
API cost & lower & higher \\
Quality claim & baseline & no statistical claim \\
Best use & stable production fallback & adaptive execution \\
\bottomrule
\end{tabularx}
\end{table}
\noindent\textbf{6.5 Early self-evolution signal (supporting).} We report the autonomous topic-tuning loop using relative, normalized, anonymized measures only (no absolute revenue). Across three self-operated accounts we compare a pre-loop baseline window with a post-loop window (Table 6); revenue is normalized per account (baseline $\equiv$ 1.0) and gains reflect both higher hit-rate and a shift toward higher-eCPM lanes. We do not claim a causally isolated effect: the comparison is observational, not an ablation, and platform/seasonal confounders are uncontrolled. Mixed outcomes (including a reversion) are reported as-is.

\begin{table}[htbp]\centering\footnotesize
\caption{Early self-evolution signal (relative, normalized, anonymized; no absolute figures).}
\begin{tabularx}{\textwidth}{l X X X X X}
\toprule
\textbf{Account} & \textbf{Baseline window} & \textbf{Post-loop window} & \textbf{Hit-rate change} & \textbf{Norm. revenue} & \textbf{Strategy update} \\
\midrule
A & Week 1--2 & Week 3--4 & +58.0\% & 1.00 $\to$ 4.23 & favored topic type 1 \\
B & Week 1--2 & Week 3--4 & +23.7\% & 1.00 $\to$ 1.86 & reduced topic type 2 \\
C & Week 1--2 & Week 3--4 & $-$2.3\% & 1.00 $\to$ 0.95 & reverted strategy \\
\bottomrule
\end{tabularx}
\end{table}
{\footnotesize\itshape Note: normalized-revenue change is larger than hit-rate change because revenue is the product of impressions and eCPM, and eCPM depends strongly on the chosen lane/topic. The self-evolution loop not only raises hit-rate but shifts topic selection toward higher-eCPM lanes, so the two gains compound. Account C is retained as a real reverted case; we report mixed outcomes as-is, not curated ones.}

\section{Convertibility Taxonomy}
The precondition for everything above---composability and the self-evolution it enables---is that a workflow can be decomposed into stages with independent, contractable boundaries. Not every workflow can. This section specifies the criteria that the convertibility-routing stage (Section 4) applies to classify and route an incoming workflow; the taxonomy is thus not a static classification but the decision logic of a stage in the harness. The single operational test: can each step be executed independently, given only a typed input, and produce a typed output?

\begin{table}[htbp]\centering\footnotesize
\caption{Diagnostic checklist.}
\begin{tabularx}{\textwidth}{l X X X}
\toprule
\textbf{Question} & \textbf{Type A} & \textbf{Type B} & \textbf{Type C} \\
\midrule
Each step has a callable function? & yes & partial & no \\
Decisions embedded in prompts? & low & high & total \\
Can a step run independently? & yes & after refactor & no \\
Input/output schema-typable? & yes & partly & no \\
Migration risk & low & medium & high \\
\bottomrule
\end{tabularx}
\end{table}
\begin{table}[htbp]\centering\footnotesize
\caption{Migration cost estimate by type.}
\begin{tabularx}{\textwidth}{l X X X}
\toprule
\textbf{Type} & \textbf{Required work} & \textbf{Cost} & \textbf{Risk} \\
\midrule
A & wrap + toolify & low & low \\
B & prompt decomposition + toolify & medium & medium \\
C & Code-First rewrite & high & high \\
\bottomrule
\end{tabularx}
\end{table}
\begin{table}[htbp]\centering\footnotesize
\caption{Worked examples.}
\begin{tabularx}{\textwidth}{l X X}
\toprule
\textbf{Workflow} & \textbf{Class} & \textbf{Reason} \\
\midrule
WeChat content automation & Type A & functions already callable \\
One-shot prompt content generation & Type C & no separable stages \\
E-commerce listing generation & Type B/A & depends on prompt/tool split \\
Customer-service agent & Type B & policy \& action logic mixed in prompts \\
\bottomrule
\end{tabularx}
\end{table}
\noindent\textbf{Failure case.} We attempted to decompose a monolithic single-prompt workflow directly into stages and failed: no resulting stage could be independently replayed or evaluated, because all judgment lived inside one prompt. The workflow had to be Code-First refactored before any stage boundary existed. A taxonomy with only positive cases reads as hindsight; this negative case is what makes the criterion a finding---convertibility must be established before, not assumed during, migration.

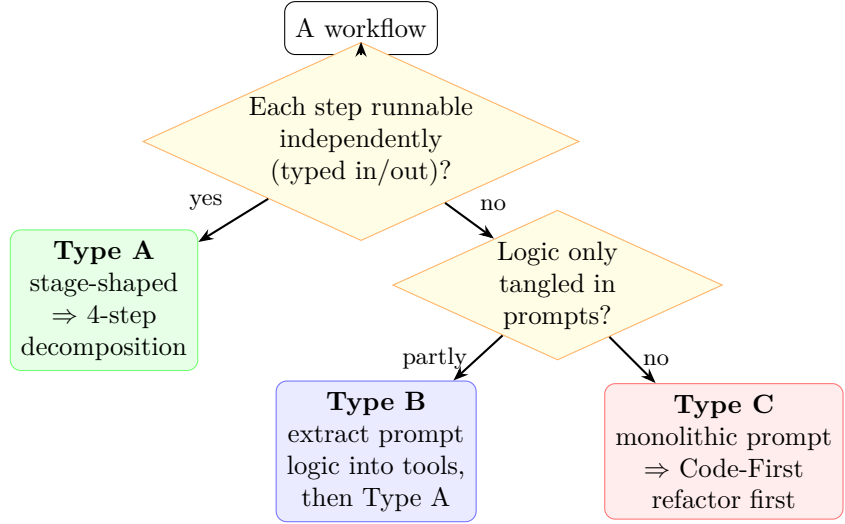
\begin{figure}[htbp]
\centering
\begin{tikzpicture}[
  font=\small,
  q/.style={draw, diamond, aspect=2.2, align=center, inner sep=1pt,
            fill=yellow!12, draw=orange!60},
  t/.style={draw, rounded corners, align=center, minimum height=7mm,
            inner sep=4pt},
  ta/.style={t, fill=green!10, draw=green!55},
  tb/.style={t, fill=blue!8, draw=blue!55},
  tc/.style={t, fill=red!7, draw=red!55},
  arr/.style={-{Stealth[length=2mm]}, thick},
]
\node[t] (wf) at (0,0) {A workflow};
\node[q] (q1) at (0,-1.5) {Each step runnable\\independently\\(typed in/out)?};
\node[ta](A)  at (-3.4,-3.6) {\textbf{Type A}\\stage-shaped\\$\Rightarrow$ 4-step\\decomposition};
\node[q] (q2) at (2.6,-3.4) {Logic only\\tangled in\\prompts?};
\node[tb](B)  at (0.2,-5.6) {\textbf{Type B}\\extract prompt\\logic into tools,\\then Type A};
\node[tc](C)  at (4.8,-5.6) {\textbf{Type C}\\monolithic prompt\\$\Rightarrow$ Code-First\\refactor first};
\draw[arr](wf)--(q1);
\draw[arr](q1)-- node[above left]{\footnotesize yes} (A);
\draw[arr](q1)-- node[above right]{\footnotesize no} (q2);
\draw[arr](q2)-- node[left]{\footnotesize partly} (B);
\draw[arr](q2)-- node[right]{\footnotesize no} (C);
\end{tikzpicture}
\caption{Convertibility decision procedure. The single test---can each step run
independently on a typed contract?---routes a workflow to Type A (decompose
mechanically), B (refactor prompts first), or C (Code-First refactor before any
stage exists).}
\label{fig:abc}
\end{figure}

\section{Related Work}
We position against four lineages rather than competing as ``yet another agent framework.'' Self-improving agents (memory/RL/policy learning)~\cite{memoryr1,agemem} operate at the parameter level on benchmarks; safety guardrails operate at the mechanism level and mostly target external prompt injection (an orthogonal threat~\cite{shieldagent,agentspec,detsec2602,layered2604} to our agent's own decision boundary); agent frameworks (ReAct, Reflexion~\cite{reflexion}, AutoGen~\cite{autogen}, Planner--Executor, Toolformer~\cite{toolformer}) are greenfield~\cite{react,planact,agentsurvey,landscape}; legacy migration (Strangler Fig, modernization surveys)~\cite{legacysurvey,legacyai} addresses traditional software, not LLM+script workflows. Our gap is the intersection none occupies: migrating a running expert workflow toward a self-evolution-ready system.

\begin{table}[htbp]\centering\footnotesize
\caption{Related-work comparison.}
\begin{tabularx}{\textwidth}{l X X X}
\toprule
\textbf{Category} & \textbf{Focus} & \textbf{Setting} & \textbf{Limitation vs this work} \\
\midrule
Agent frameworks & planning / tool use & greenfield & not legacy migration \\
Safety guardrails & prevent unsafe actions & mechanism-level & not workflow decomposition \\
Self-improving agents & memory / RL / policy & benchmark / model-level & not production workflow strategy \\
Legacy migration & software modernization & traditional software & not LLM+script workflows \\
This work & reversible migration to composable harness & production workflow & single-case limitation \\
\bottomrule
\end{tabularx}
\end{table}
\section{Discussion and Threats to Validity}
Cross-vertical reuse (early). A second vertical (e-commerce) is being instantiated on the same harness. As quantitative reuse data are not yet available, we report a conceptual reuse mapping only (Table 11), and place reuse rates, new-tool counts, and time-to-first-recipe as ongoing work. We deliberately do not report fabricated LOC/time numbers for the second vertical.

\begin{table}[htbp]\centering\footnotesize
\caption{Cross-vertical conceptual reuse (e-commerce, ongoing).}
\begin{tabularx}{\textwidth}{l X X X}
\toprule
\textbf{Component} & \textbf{WeChat} & \textbf{E-commerce} & \textbf{Reuse (design)} \\
\midrule
recipe loader & yes & yes & full \\
trace store & yes & yes & full \\
agent loop & yes & yes & full \\
safety gate abstraction & yes & yes & partial \\
selection stage & topic selection & product discovery & conceptual \\
publishing stage & draft save & listing publication & adapted \\
\bottomrule
\end{tabularx}
\end{table}
Threats to validity. Single case (no statistical-significance claims; generalizability open). Self-evolution scope (an early, non-causal strategy-level signal, not a validated learning result). Model dependency (a relay API proxying Claude-class models; cross-model consistency unvalidated). Threat-model scope (adversarial tests cover the agent's own decision boundary only; external prompt injection is orthogonal and uncovered). Planner layer (higher-level scheduling remains largely deterministic).

\section{Conclusion}
We report a reversible, auditable, low-risk path that turns frozen expert workflows into composable, self-evolution-ready harnesses, validated on a production workflow by migration cost, reversibility, stage contracts, audit trace, deterministic safety invariants, and an early outcome-feedback signal.

The practical bottleneck in agentifying expert workflows is not model capability alone, but convertibility: whether business judgment can be separated into typed, independently runnable stages. This paper provides the migration path and the diagnostic taxonomy for that conversion.

\end{document}